\documentclass[12pt]{article}

\usepackage{amsmath,amssymb,mathrsfs}
\usepackage{color}
\usepackage{eso-pic} 
\usepackage{lscape} 
\usepackage{multirow}
\usepackage{rotating}
\usepackage{float}
\begin{document}

\thispagestyle{empty}
%\vspace*{-3.5cm}
%\begin{flushright}
%CDF/ANAL/BOTTOM/CDFR/8857\\
%Version 0.0\\
%\today
%\end{flushright}

%\vspace{0.25cm}

\begin{center}
  \begin{Large}
    {\bf
      A Statistical Prescription to Estimate Properly Normalized Distributions of Different Particle Species
    }
  \end{Large}
\end{center}

%\vspace{0.2cm}

\begin{center}
M.~Casarsa$^{\,a}$\footnote{casarsa@fnal.gov}
P.~Catastini$^{\,b}$\footnote{pierluigi.catastini@pi.infn.it}
G.~Punzi$^{\,c}$\footnote{giovanni.punzi@pi.infn.it}
L.~Ristori$^{\,d}$\footnote{luciano@fnal.gov}
\vspace{0.25cm}

\emph{\small{$^a$ Fermi National Accelerator Laboratory, Batavia, USA}} \\
\emph{\small{$^b$ Fermi National Accelerator Laboratory, Batavia, USA}} \\
\emph{\small{$^c$ Universit\`a di Pisa and INFN Sez. Pisa, Pisa, Italy}} \\
\emph{\small{$^d$ INFN Sez. Pisa, Pisa, Italy}}
\end{center}

\vspace{0.25cm} 

\begin{abstract} 
We describe a statistical  method  to avoid
biased estimation of the  content  of different  particle species. We consider the case when 
the particle identification information strongly
depends on some kinematical variables, whose distributions are unknown and different 
for each particles species.
We show that the proposed procedure provides properly normalized and completely data-driven
estimation of the unknown distributions without any a priori assumption on
their functional form. Moreover, we demonstrate that the method can
be generalized to any kinematical distribution of the particles.
\end{abstract}

% ======================================================================
\section{Introduction}
\label{sec:intro}
% ======================================================================
%
The  estimation of  the particle species content  in a
sample  of reconstructed  tracks is  a recurrent  problem  in Particle
Physics.  To this purpose,  different experimental techniques are used
to obtain  information about the  particle type; typical  examples are
the measurement of the particle Time-of-Flight (\emph{ToF\/}) from the
production  vertex to  a given  position inside  the detector,  or the
measurement  of  the particle  energy  loss  per  unit length  of  the
traveled  path  due to  the  interaction  with  the detector  material
($dE/dx$).   \\ The information provided by these techniques is related to the particle type but typically
 depends also on the track momentum.\\ It  is   common  practice  to  include  the  particle
identification   ({\it   PID\/})  information   in   a  Maximum
Likelihood  (ML) fit  in  order to  estimate  the particle  species
content  of the  sample. On the other hand, the strong momentum dependence
of the separation power between different particles may lead to strongly 
biased results, if not properly treated in the ML fit. \\
To  be more  specific,  let's  consider a  mixture  of known  particle
species,  for   example  pions  ($\pi$),   kaons  (\emph{K}),  protons
(\emph{p}),  and  electrons  (\emph{e}), and assume that the \emph{PID} 
information  is provided by the $dE/dx$ measurement in a drift
chamber.   Using  the   separating  power  provided  by  the
\emph{PID},  we  want to  estimate the unknown fractions $f_{\pi}$, 
$f_{K}$, $f_{p}$, and $f_{e}$ of each particle  type contained in  the 
sample by means of a ML fit. 
Our observables are the measured $dE/dx$ response (that we will indicate 
as \emph{x}) and  the momentum of the particle \emph{p}. 
We then label as $t_{j}$ the particle hypothesis and  the conditional  
probability  density function  of $x_{i}$ for track $i$,  given $p_{i}$   and $t_{j}$, will  
be  indicated  as   ${\mathcal   P}(x_{i} | p_{i},\,t_{j})$.\\   
Finally, the likelihood function is expressed as:
\begin{eqnarray}                                   \label{eq:likeright}
\lefteqn{L(f_{j})  =  \prod\limits_{i}(\sum\limits_{j=\pi,K,p,e}   f_{j}  {\mathcal
P}(x_{i},\,p_{i}\:|\, t_{j}))  {}} \nonumber \\
 & & {} =  \prod\limits_{i}(\sum\limits_{j=\pi,K,p,e}  f_{j} {\mathcal
P}(x_{i}  | p_{i},\, t_{j})  \times {\mathcal  P}(p_{i} |
t_{j}) \quad ) \ ,
\end{eqnarray} 
where $i$ is the track index, with the additional condition:
\begin{equation} \sum\limits_{j=\pi,K,p,e} f_{j} = 1 \ .
%\label{eq:condition}
\end{equation}
In  practice, we  often  have poor
information on the distributions  of  the additional  observables
(${\mathcal P}(p_{i}  | t_{j})$ in  our example).  Sometimes  they are
completely  unknown.  This  is  the case,  for  example, of  particles
produced during the hadronization of $B$ mesons\footnote{This work was
motivated by the effort of understanding
the properties  of particles produced during  $B$ mesons hadronization,
that represents  a  major  issue  in  the  development  of  flavor  tagging
algorithms  like  the Same  Side  Kaon  Tagging \cite{mixsskt}.}:  the
momentum distribution of each particle type is unknown and the correct
likelihood  function  as  defined  in (\ref{eq:likeright})  cannot  be
constructed.\\ 
Note that using the conditional likelihood function:
\begin{equation}    L(f_{j})=\prod\limits_{i}(\sum\limits_{j=\pi,K,p,e}
f_{j} {\mathcal P}(x_{i} | p_{i},\,t_{j}))\ 
\label{eq:likewrong}
\end{equation}
may  lead to   strongly  biased results,  if  our additional  variable,  the  momentum in our example, has  different
distributions   for  different  particle   types since the ${\mathcal P}(p_{i}  | t_{j})$
 term  cannot be factorized in (\ref{eq:likeright}).  As   discussed  in 
 \cite{punzi}  and shown  in \cite{pierlugio} specifically for the particle species estimation,  
whenever  the templates used in a multi-component fit  depend on additional observables, it is
necessary to  use the  complete likelihood expression,  and explicitly
include  the  probability distributions  of  all  observables. In  our
example this implies that we need to include
the momentum distributions of each particle type  in the likelihood. 
The crucial question is how to avoid strong bias in the particle fraction estimation when 
the momentum distributions of each particle type are unknown.\\
In  \cite{pierlugio}  it  was shown  that  a  possible
solution to this  problem is to use a series  expansion of the unknown
distributions;  the  Fourier  coefficients  of  the  series  are  free
parameters determined  by the fit.\\ 
In this paper we propose a different strategy, based on the  idea that 
if the fit is  performed in sufficiently small momentum intervals, the 
bias due to the use of the likelihood function (\ref{eq:likewrong}) is 
small  and it  goes to  zero as  the momentum interval width decreases. 
In  Sec.~\ref{sec:likelihood} we present a simple procedure to perform the fit and show  that  we can  
extract  the unknown  momentum distributions of each  particle type in 
a completely  data driven mode without  any   {\it  a  priori\/}  
assumption   on  the  corresponding functional  forms.\\  
In Sec.~\ref{sec:other}, given the results of our fitting procedure, we describe a 
powerful  technique to extract the distribution of  any kinematical 
variable for any particle type, with the proper  normalization.  
Finally, in Sec.~\ref{sec:pythiamc} some applications of the method
to Pythia  ~\cite{Pythia} Monte Carlo  samples simulated with  the CDF
detector ~\cite{cdfdet} are shown.
%
%======================================================================
\section{Estimation of Particle Fractions and Momentum Distributions}
\label{sec:likelihood}  
%======================================================================
% 
As anticipated above, we  restrict ourselves to  the particles  contained 
in  small momentum intervals of equal width $\Delta  p$. In  each bin we  
estimate the  particle content using a ML fit by observing  that if $\Delta p$ 
is sufficiently small, the   bias  introduced   by  using   the  conditional   
likelihood  of (\ref{eq:likewrong}) is  negligible. 
As will be evident in the following, we find more convenient to use an 
Extended Likelihood (EL) fit (see \cite{cowan} for a definition).
In each momentum bin $m$ the Extended Likelihood function takes 
the form:
\begin{eqnarray}   \log(L_{m})    =   \mathcal{L}_{m}   =\sum\limits_{i=0}^{N_{m}}
\log(\sum\limits_{j=1}^{M}  N_{j,m}\,{\mathcal  P}_{j}(x_{i}| m, t_{j}))-       
N_{m}       \\      \nonumber{=       \sum\limits_{i=0}^{N_{m}}
\log(\sum\limits_{j=1}^{M}   N_{j,m}\,{\mathcal  P}_{j}(x_{i}|m,
t_{j}))- \sum\limits_{j=1}^{M} N_{j,m}}\ ,
\label{eq:likelihoodform}
\end{eqnarray}
where $N_m$  is the total number  of particles in the $m$-th momentum
bin, $M$ is the number of different particle species, $N_{j,m}$
is the number of particles of  type $j$, $x_{i}$ is the {\it PID\/} 
response for the $m$-th momentum bin and ${\mathcal P}_{j}(x_{i}| m,
t_{j})$ is the conditional  probability density function associated to
$x$ for the  particle  type $t_{j}$ in the $m$-th momentum  bin . 
In  the rest of  the paper we  will simplify the  notation by
setting ${\mathcal  P}_{j}(x_{i}) = {\mathcal  P}_{j}(x_{i}| m, t_{j})$.
The EL must be maximized with respect to the free parameters $N_{j,m}$.\\ 
We notice  that, given  its particular  form,  the first  derivative 
of our EL function can be evaluated analytically as:
\begin{equation}
  \frac{\partial{\mathcal{L}_{m}}}{\partial{N_{j,m}}}=
\sum\limits_{i=1}^{N_{m}}\frac{{\mathcal P}_j(x_{i})}
{\sum\limits_{k=1}^{M} N_{k,m } {\mathcal P}_k(x_{i})} - 1 \ .
\label{eq:derive}
\end{equation}
The critical points of  (\ref{eq:likelihoodform})  can be obtained using an iterative algorithm.
 In  particular, starting  from a first guess on our parameters,  $N_{j,m}^{0}$, we can define an
iterative procedure, a \emph{Picard iteration}~\cite{ortega}, of the
form:
\begin{equation}  N_{j,m}^{n}  =  \sum\limits_{i=1}^{N_{m}}\frac{N_{j,m}^{n-1}
{\mathcal      P}_j(x_{i})}{\sum\limits_{k=1}^{M} N_{k,m}^{n-1} {\mathcal P}_k(x_{i})} \ ,
\label{eq:itercl}
\end{equation}
where at every step $n$  of the iteration, we estimate the $N_{j}^{n}$
that will be  used as input values for the  step $n+1$.  The iteration
converges to the roots  of (\ref{eq:derive}).  Similarly, we can write
the    analytical   expression   of    the   second    derivative   of
(\ref{eq:likelihoodform})  and estimate the  statistical error  of our
parameters   from   the  inverse   of  the   covariance  matrix
$V_{jl,m}^{-1}$ of the fit in the $m$-th momentum bin as:
\begin{equation}     V_{jl, m}^{-1}=\frac{\partial (-\mathcal{L}_{m})}{\partial
N_{j,m}\partial{N_{l,m}}}   =\sum\limits_{i=0}^{N_{m}}
\frac{{\mathcal P}_j(x_{i}){\mathcal P}_l(x_{i})}{\biggr(\sum\limits_{k=1}^{M}     
N_{k,m} {\mathcal P}_k(x_{i})\biggr)^{2}} \ .
\label{eq:covacl}
\end{equation}
Our iterative procedure is equivalent to the \emph{Channel Likelihood}
method introduced in ~\cite{chanlike}. It can be easily shown that the
Channel Likelihood method  is a maximization of an  EL function of the
form (\ref{eq:likelihoodform}) using an  iterative scheme similar to the one
we  propose.\\ This  approach  has several  advantages: no  functional
shape is  assumed for the  momentum spectra, the method,  provided the
 ${\mathcal  P}_{j}(x) $ templates, is  completely data driven; all the  fits can be
performed in parallel,  that is the iteration can be  done in all 
momentum bins at the same time; the algorithm is very fast and stable 
respect to the initial guess on the $N_{j,m}$.\\
Once we obtain convergence of  the iterative process, we can write the
fraction of each particle type as:
\begin{equation}        f_{j}        =        \frac{N_{j}}{N}        =
\frac{1}{N}\sum\limits_{m} N_{j,m} \ .
\label{eq:fractions}
\end{equation}
Observing  that   each  bin  content  is   fitted  independently,  the
corresponding statistical uncertainty is:
\begin{equation} \sigma(f_{j})  = \frac{\sqrt{\sigma^{2}(N_{j})}}{N} =
\frac{1}{N}\sqrt{\sum\limits_{m}  \sigma^{2}(N_{j,m})} \
.
\label{eq:sigmafractions}
\end{equation}
Finally, since we  are performing the fit in each  momentum bin, the  arrays  of
$N_{j,m}$  can   be  interpreted  as  \emph{histograms},  that
reasonably  approximate  the   true  momentum  distributions  of  each
particle  type. In this way, we  obtained an  unbiased estimation  of  the particle
composition, although the ${\mathcal  P}(p_{i} | t_{j})$ were unknown,
and,  at the  same time,  a  reasonable estimation  of the  ${\mathcal
P}(p_{i} | t_{j})$ themselves.\\
%
%======================================================================
%\subsection{A Monte Carlo Study}
%\label{sec:mc}                                                        
%======================================================================
%
\begin{figure}[!t]
\begin{center}
 \includegraphics[width=5.5in]{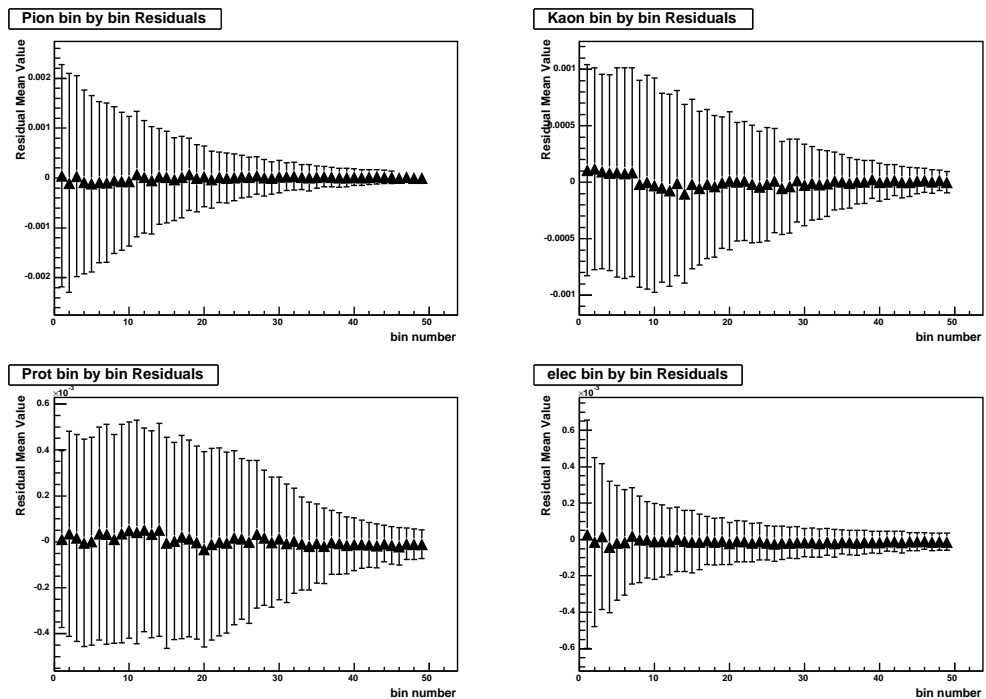}
\caption{Mean and width  of the distribution of the  difference of the
true and measured fractions in each momentum bin.}
\label{fig:resbin}
\end{center}
\end{figure}
We test our method on parametric Monte Carlo samples. 
    Each    sample     consists     of    20,000
particles.  Different  momentum  spectra  are used  to  generate  each
particle type and the corresponding fractions are fixed at: $f_{\pi} = 0.74$, 
$f_{K} = 0.17$, $f_{P}=0.07$, and $f_{e}=0.02 \ $.
%
%\begin{equation} f_{\pi} = 0.74, \quad f_{K} = 0.17, \quad f_{P}=0.07,
%\quad f_{e}=0.02 \ .
%\label{eq:fractionstoycl}
%\end{equation}
%
We divide the sample in  50 momentum bins from 0.45 GeV/$c$ up to
5  GeV/$c$ and a  EL fit  is performed  in each  momentum bin.   If no
particle falls in a given  bin, the corresponding fit is not performed
for obvious reasons.   We repeat the fit on  500 parametric samples to
extract  the   residual  distributions   of  the  estimators.  No
significant bias is observed as summarized in Tab.~\ref{tab:toy2}. \\ Another interesting test 
 is the residual distribution of the estimators   in  each   momentum  bin   for  the   500   samples.  
Fig.~\ref{fig:resbin} shows the residuals of the four estimators as  a function of
the bin  number (i.e.  momentum).  In each plot,  each point
represents the mean value, while the error bar represents the width of
the distribution of  the residual.  We observe no  significant bias. 
Examples of the unknown momentum distributions obtained from the fit are shown in Fig.~\ref{fig:pt}.\\ 
A very good agreement between the
true distributions (filled histograms)  and the fit (dots) is observed
for all particle types. 
\begin{table}[htbp] \centering
  \begin{tabular}{|c|c|c|c|c|}    \hline     &    $f_{\pi}^{true}    -
f_{\pi}^{est.}$  &  $ f_{k}^{true} -  f_{k}^{est.} $ &  
$f_{p}^{true} - f_{p}^{est.}$ & $f_{e}^{true}  - f_{e}^{est.}$ \\ \hline  EL &  
$ (0.2 \pm 2) \times 10^{-4} $ & $(2 \pm 2)\times 10^{-4}$ &  $ (3 \pm  1)\times 10^{-4} $ &  
$ (5 \pm 1)\times 10^{-4} $ \\ \hline
  \end{tabular}
  \caption{{\small  Mean  values  of  the  residuals  of  the  particle
fractions (integrated on all momentum bins) estimated in 500 parametric samples.}
    \label{tab:toy2}}
\end{table}
%
%
%
%======================================================================
\section{Extracting distributions of other quantities}
\label{sec:other}                                                    
%======================================================================
%
%\begin{figure}[!t]
%\begin{center}
% \includegraphics[width=5.5in]{scheme6.eps}
%\caption{Overlay of the  parametric Monte Carlo momentum distributions
%(filled histograms) and the ones resulting form the EL fits (dots).}
%\label{fig:fitpt}
%\end{center}
%\end{figure}
%
Besides estimating the momentum spectra for the different particle  
species, we are typically also interested in obtaining distributions of 
additional relevant kinematical variables. We achieve this by combining 
our fitting procedure with a technique known as {\it sPlot\/}~\cite{splot}, which
allows to estimate the composition of a mixture of several components 
and their covariance matrix by means of an Extended Maximum-Likelihood
fit.\\
The main requirement for  the {\it sPlot\/} technique is that the 
kinematical variables to be plotted be uncorrelated with the discriminating 
variables of the fit.
The output is represented by an histogram, called {\it sPlot\/}, of  the
relevant variables.\\ 
In  our problem, this  means that the  variables we want to plot must  
not be correlated  with the \emph{pdf}'s  that describe the {\it PID\/}
response. However,  any   dependence  of  the  \emph{pdf}'s  on
kinematical variables is  {\it frozen\/} by our strategy  to perform a
separate fit  in each  momentum bin: the  correlation of  any variable
with  the momentum,  if any,  can be  neglected inside  a sufficiently
small bin. \\ 
For each  momentum bin, our fit provides  an estimate of
the  particle  content and  the  corresponding  covariance matrix.  It
follows  that for  each bin  we can  produce the  \emph{sPlot}  of any
kinematical   variable.   \\   Suppose   we  want   to   produce   the
$sPlot_{j,m}(y)$ of a given  variable $y$ for the particle type
$j$,  corresponding to the EL fit  performed in  the $m$-th  bin. As
explained in \cite{splot},  we have to properly combine  the result of
the fit in the $m$-th bin with the corresponding covariance matrix
to define an event by event weight called the s-weight, $sw_{ij,m}(x)$.  
We  then have  to fill an  histogram of $y$  where each event $i$, 
falling in the $m$-th bin, is weighted according to the s-weight:
\begin{equation}          sw_{ij,m}(x)         =
\frac{\sum\limits_{l=1}^{M}V_{jl,m}      \:      {\mathcal
P}_l(x_{i})}{\sum\limits_{k=1}^{M}  N_{k,m}  \:{\mathcal
P}_k(x_{i})} \ ,
\label{eq:sweight}
\end{equation}
where  $N_{k,m}$  are  obtained by  performing  the  iteration
(\ref{eq:itercl})  and  $V_{jl,m}$ is  the  covariance
matrix (\ref{eq:covacl}) corresponding to the fit performed in the 
$m$-th bin. Applying the  above procedure in  each momentum interval,  
we obtain an array of $sPlot_{j,m}(y)$ histograms.\\ 
We then exploit the additive property of the \emph{sPlot}'s to add
together all the $sPlot_{j,m}(y)$ corresponding to a given variable, 
one for each momentum bin;
the resulting \emph{sPlot} represents the distribution of the given 
variable $y$ for the whole momentum range:
\begin{equation}     sPlot_{j}(y)     =     \sum\limits_{m}
sPlot_{j,m}(y) \ .
\label{eq:sp}
\end{equation}
Thanks  to the  combination of  the \emph{sPlot}  and our  strategy of
performing several fits  in small momentum bins, we are  able to obtain the
distribution  of  any  kinematical  variable  of  each  particle  type
belonging to our initial sample.  \\
The method can be summarized in a three-step procedure as follows:
\begin{enumerate}
\item Fit  the particle species content in  several momentum intervals
by means of the Extended Likelihood method via an iterative scheme.
\item In each momentum bin,  using  the  results of  the fits, produce  the
\emph{sPlot}'s of  any additional variable  you are interested  in for
all the particle species.
\item   Separately  for  each   particle  type,   add  the   array  of
\emph{sPlot}'s of a given variable  to extract the distribution of the
variable on the whole data sample.
\end{enumerate}
%
%======================================================================
\section{Monte Carlo Study of the Complete Procedure}
\label{sec:pythiamc}                                                  
\begin{figure}[!t] \centering
  \includegraphics[width=5.0in]{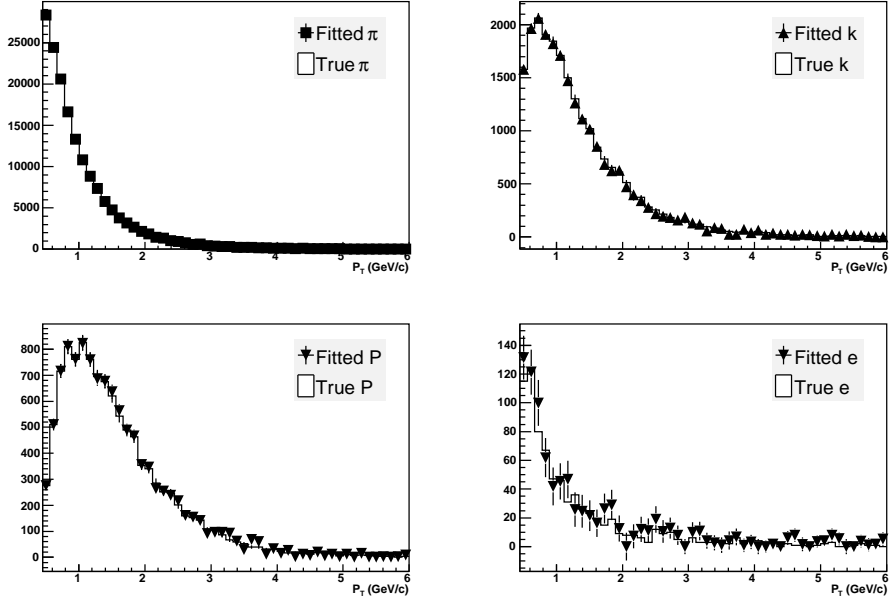}
  \caption{Overlay of the MC $p_{T}$ distributions (filled histograms)
and the result from the fit (dots).
  \label{fig:pt}}
\end{figure}

\begin{figure}[!b] \centering
  \includegraphics[width=5.0in]{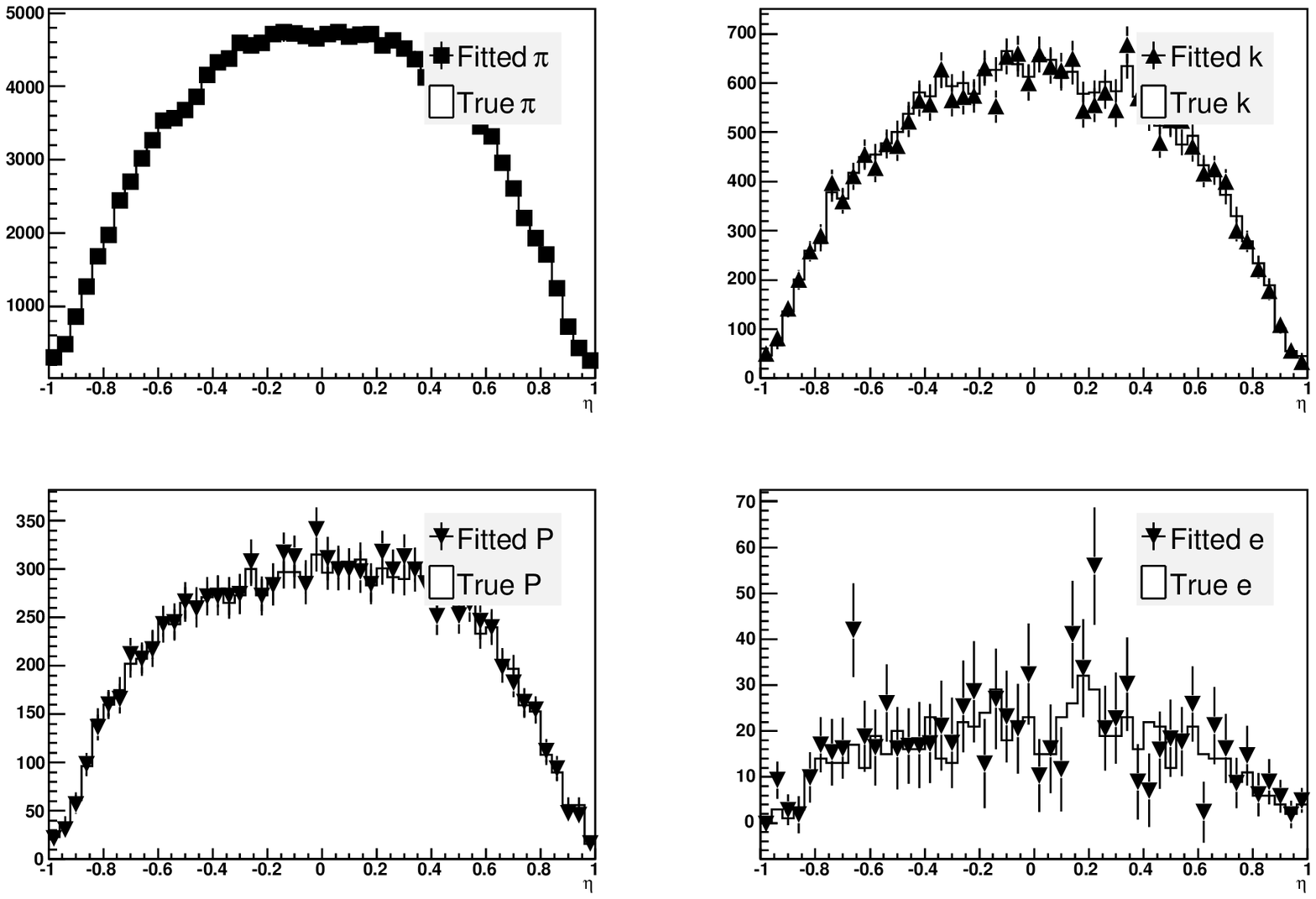}
  \caption{Pythia   MC   $\eta$    distributions   overlaid   to   the
corresponding data {\it sPlot}.
    \label{fig:eta}}
\end{figure}

\begin{figure}[!t] \centering
  \includegraphics[width=5.0in]{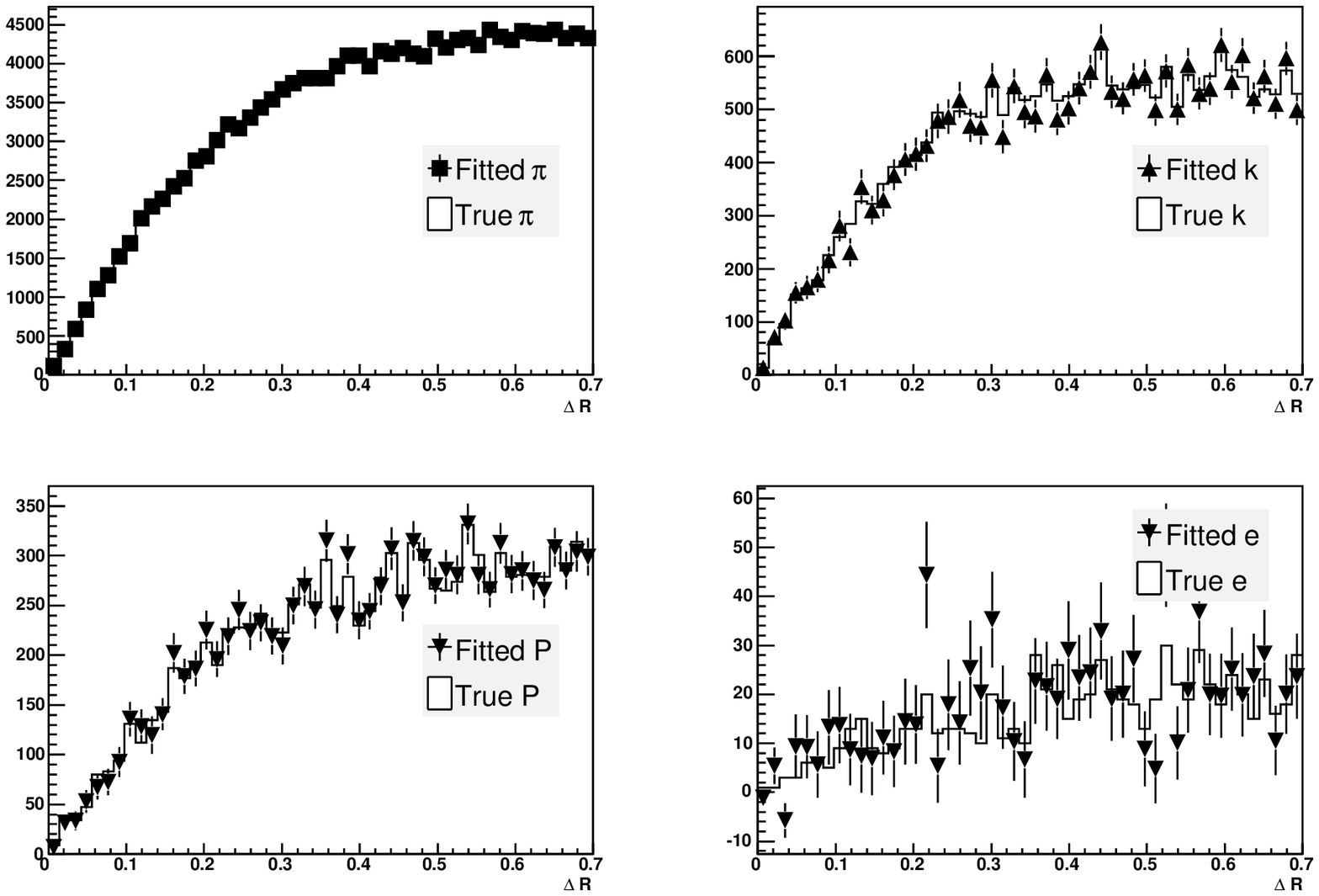}
  \caption{Pythia  MC   $\Delta  R$  distributions   overlaid  to  the
corresponding data {\it sPlot\/}.
    \label{fig:deltaR}}
\end{figure}

\begin{figure}[!b] \centering
  \includegraphics[width=5.0in]{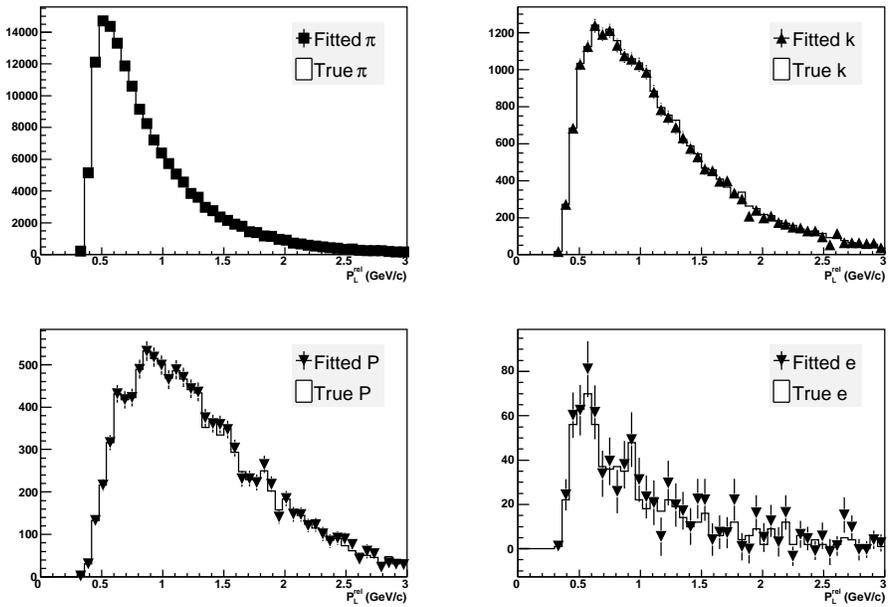}
  \caption{Pythia  MC  $p_{L}^{rel}$  distributions  overlaid  to  the
corresponding data {\it sPlot\/}.
    \label{fig:plrel}}
\end{figure}

%======================================================================
%
%\begin{figure}[!tbp] \centering
%  \includegraphics[width=3.5in]{scheme9.eps}
%  \caption{{\small  Overlay of  Pythia $\eta$  distributions  for each
%particle type.}
%    \label{fig:etaconf}}
%\end{figure}
%
To test the whole procedure in a realistic case, we use a Pythia Monte 
Carlo~\cite{Pythia} sample  of  $B^{-}\!\rightarrow\!    D^{0}\pi^{-}$
events processed with the CDF  detector simulation plus a parametric 
simulation of $dE/dx$ and $ToF$. Such  a  sample provides a reasonable
description of  all the kinematical variables associated  with a given
particle and the correlations among them.  We then fit the composition
of   the  particles   produced  in   the  vicinity   of   the  $B^{-}$
meson\footnote{This  is  a  typical  problem  encountered  during  the
development and  the calibration of flavor tagging  algorithms for the
$B$ meson.}.\\
The  \emph{ToF} response for a particle depends on  two observables: the 
momentum  and  the  length  traveled  by  the  particle  (also  called
arc-length). Consequently, the  use of  the \emph{ToF}  requires, in
principle,  a two-dimensional  binning in  momentum  and arc-length. 
However, given the cylindrical geometry of the  CDF detector,  momentum
 and arc-length can be  replaced by the momentum component in the 
transverse  plane $p_{T}$ and the pseudorapidity $\eta$\footnote{The 
pseudorapidity of a track is defined as $\eta = -\log(\tan(\frac{\theta}{2}))$, 
where $\theta$ is the polar angle.}  with no loss of information. 
%In Fig.~\ref{fig:etaconf} we show the $\eta$  distributions   for  different
%particle types: since no significant difference is observed for the four
%particle species, this allows to  factorize the $\eta$ distribution and 
%eliminate it from the likelihood.
%
Tab.~\ref{tab:fracpythia} reports the  total fraction of  each  particle  
type   and  Fig.~\ref{fig:pt}  shows  the  $p_{T}$ distributions resulting 
from the fit.\\
Combining the  particle content and the  covariance matrix estimated
in each $p_{T}$ bin, we are able to produce the {\it sPlot\/} of other
kinematical   variables   for   the   whole  Monte   Carlo   sample.
Some examples are shown in Figs.~\ref{fig:eta}-\ref{fig:plrel}: track $\eta$,
the distance  $\Delta R  = \sqrt{\Delta \phi^{2}  + \Delta{\eta}^{2}}$
between the particle  and the $B$ meson flight  directions, 
the longitudinal  component  $p_{L}^{rel}$ 
of the  particle momentum with respect to  the $B$ meson
flight direction (this variable  is often used  in the  $B$ flavor
tagging algorithms).\\  A very good  agreement is observed  between the
true  distributions and  the estimated ones, even  in  those cases
where the plotted variable is strongly correlated to the $p_{T}$.

\begin{table}[htbp] \centering
  \begin{tabular}{|c|c|c|c|c|} \hline  & $\pi$  & $K$ &  $p$ &  $e$ \\
\hline  Generated  &  $0.828$  &  $0.112$ &  $0.0557$  &  $0.0039$  \\
Estimated  & $0.829  \pm  0.002$ &  $0.111  \pm 0.001$  & $0.0557  \pm
0.0006$ & $0.0044 \pm 0.0003$ \\ \hline
  \end{tabular}
  \caption{{\small  Comparison  between   the  true  particle  content
in  the  Pythia  Monte  Carlo sample (Generated)   and  the  fractions
resulting from the fits (Estimated).}
    \label{tab:fracpythia}}
\end{table}
\clearpage
 %\clearpage
%\begin{figure}[!hb] \centering
%  \includegraphics[width=5.0in]{scheme13.eps}
%  \caption{Pythia MC $M(B\pi) -  M(B) - M(\pi)$ distributions overlaid
%to the corresponding data {\it sPlot\/}.
%    \label{fig:qbd}}
%\end{figure}
%
 %\clearpage
%\begin{figure}[!hb] \centering
%  \includegraphics[width=5.0in]{scheme15.eps}
%  \caption{Pythia  MC  $P_{T}^{rel}$  distributions  overlaid  to  the
%corresponding data {\it sPlot\/}.
%    \label{fig:ptrel}}
%\end{figure}                     
%

%======================================================================
\section{Conclusions and outlook}
\label{sec:conc}                                                      
%======================================================================
%
We  have   presented  a  method  to  estimate   the  distributions  of
kinematical variables of  different particle species using information
from {\it PID\/} detectors. This method  is shown to work very well on
Monte  Carlo  data  allowing   the  determination  of  the  fractional
composition  of  a mixed  sample  of  particle  types with  remarkable
precision.  We  use a likelihood function that  correctly contains the
$pdf$'s of all  the relevant event observables and  therefore avoids a
common mistake leading to  strongly biased estimations.  This approach
looks  very  promising  for  the  determination  of  distributions  of
different  types of  particles produced,  for example,  in conjunction
with a $B$ meson in  a completely data-driven mode, without making any
prior assumption on their functional forms.
%
%======================================================================
\section{Acknowledgments}                                            
%======================================================================

We would like  to thank the CDF Collaboration  for providing the Monte
Carlo samples used  in this paper and for  the very useful discussions
and suggestions.
%
%\clearpage                       

%------------------------------------

%

\clearpage

\end{document}